# Role of gallium diffusion in the formation of a magnetically dead layer at $Y_3Fe_5O_{12}$ / $Gd_3Ga_5O_{12}$ epitaxial interface

S. M.Suturin[1*], A. M. Korovin[1], V. E. Bursian[1], L. V. Lutsev[1], V. Bourobina[1], N. L. Yakovlev[2], M. Montecchi[3], L. Pasquali[3,4,5], V. Ukleev[6], A. Vorobiev[7], A. Devishvili[8] and N. S. Sokolov[1]

[1]Ioffe Institute, 26 Polytechnicheskaya street, St. Petersburg 194021, Russia

[2]Institute of Materials Research and Engineering, Agency for Science Technology and Research (A*STAR), 138634 Singapore

[3]Engineering Department, "E. Ferrari" University of Modena e Reggio Emilia, , Via Vigolese 905 – 41125 Modena, Italy

[4]IOM-CNR Institute, Area Science Park, SS 14 Km, 163.5 – 34149 Basovizza, Trieste, Italy

[5]Department of Physics, University of Johannesburg, PO Box 524, Auckland Park, 2006, South Africa

[6]Laboratory for Neutron Scattering and Imaging (LNS), Paul Scherrer Institute (PSI), CH-5232, Villigen, Switzerland

[7]Department of Physics and Astronomy, Uppsala University, Box 516, 751 20, Uppsala, Sweden

[8]Department of Physical Chemistry, Lund University, Box 124, SE-22100 Lund, Sweden

* Correspondence e-mail: suturin@mail.ioffe.ru

We have clarified the origin of magnetically dead interface layer formed in yttrium iron garnet (YIG) films grown at above 700°C onto gadolinium gallium garnet (GGG) substrate by means of laser molecular beam epitaxy. The diffusion-assisted formation of a Ga-rich region at the YIG / GGG interface is demonstrated by means of composition depth profiling performed by X-ray photoelectron spectroscopy, secondary ion mass spectroscopy and X-ray and neutron reflectometry. Our finding is in sharp contrast to the earlier expressed assumption that Gd acts as a migrant element in the YIG/GGG system. We further correlate the presence of Ga-rich transition layer with considerable quenching of ferromagnetic resonance and spin wave propagation in thin YIG films. Finally, we clarify the origin of the enigmatic low-density overlayer that is often observed in neutron and X-ray reflectometry studies of the YIG / GGG epitaxial system.

## I. INTRODUCTION

The intense interest to nanometer-scale epitaxial films of yttrium iron garnet ($Y_3Fe_5O_{12}$, YIG) is supported by potential applications in magnonic devices [1–3], exploiting the idea of data transfer via spin waves (SW) [4]. Magnonic applications are based on nanostructures, where SW can propagate with reduced loss over distances up to millimeters. Extremely low Gilbert damping parameter $\alpha$ =

$3.0\times10^{-5}$ and the narrowest ferromagnetic resonance (FMR) line width of $\Delta H = 0.2$ Oe of the single-crystalline YIG, make it one of the best materials in the field. Due to the absence of the three-magnon scattering [5], the spin wave damping is expected to be significantly lower in YIG ultrathin films with thickness ranging from few nanometers to few tens of nanometers. E.g. it was shown recently in Ref. [6] that SW damping in a 10 nm epitaxial YIG layer can be as low as $\alpha=3.6\cdot10^{-5}$ approaching the bulk value obtained for YIG single crystals grown by Czochralski method. Various deposition techniques including laser molecular beam epitaxy (LMBE) have been used [6–14] during the past years to grow high-quality YIG films onto the gadolinium gallium garnet ($Gd_3Ga_5O_{12}$, GGG) substrates. Despite the fact that GGG is very well lattice matched to YIG ($\Delta a/a=6\cdot10^{-4}$), it was claimed in a number of studies that the crystal structure and magnetic properties of YIG nanolayers can be quite different from the bulk. Particularly, the (111) interlayer spacing in films is often significantly larger (by 1-1.5%) than in bulk YIG due to rhombohedral distortions [13,14]. This can be caused by stoichiometry deviations due to oxygen and iron vacancies, gallium or gadolinium diffusion from the substrate, etc. The magnetooptical studies of YIG/GGG nanoheterostructures reveal a modified magnetic structure of the interface region [15]. X-ray reflectivity measurements [16] confirm presence of a few nm interface layer with a reduced density and magnetization. There exists a single polarized neutron reflectometry (PNR) study [16] showing that the interface region is paramagnetic at room temperature but becomes magnetic at 5 K and couples anti-parallel to the rest of the YIG film. Although some considerations are given therein that the interface region consists of Gd doped YIG, there is no direct evidence that the migrant element is not Ga. Moreover, scanning electron microscopy (SEM) and electron energy loss spectroscopy (EELS) have shown [16] that the interface is chemically-diffused and both Ga and Gd penetrate into the YIG film. Similarly a 5 nm thick interdiffusion region with almost zero magnetic moment was detected by PNR [17]. Energy dispersive X-ray spectroscopy (EDX) studies of YIG / GGG layers grown at 700°C [18] were interpreted in terms of a symmetrical inter-penetration of Ga, Gd, Fe, Y rather than an interdiffusion of specific elements. No asymmetrical interdiffusion was observed in YIG films grown by liquid phase epitaxy [19,20]. One can expect that this is because the growth rate in LPE is 10-100 times (micron per min) higher than in Laser MBE (10 nm / min) so the atoms cannot propagate far by diffusion, at least in the films of comparable thickness. The other reason for appearing of excess Ga or Gd at the interface in the YIG/GGG films grown by Laser MBE could be some resputtering of the substrate by energetic plasma.

In the present work we investigated in detail the YIG / GGG epitaxial layers grown at 700-

1000°C by laser MBE paying particular attention to the properties of the interface region. We studied the correlations between crystal structure, chemical composition and magnetic characteristics of thin YIG / GGG layers. We demonstrate drastic quenching of ferromagnetic resonance and spin wave propagation in ultrathin YIG films, correlating it to the structural data obtained by composition depth profiling. Secondary ion mass spectroscopy, X-ray photoelectron spectroscopy, atomic force microscopy, X-ray and neutron reflectometry are applied to demonstrate that a Ga-rich layer is formed at the bottom of the YIG film during high temperature epitaxial growth. The direct observation of Ga diffusion into the YIG film is in contrast to the earlier works claiming that the migrant element is Gd. The origin of the thin low density layer residing on top of the YIG layer is also explained. The presented results are specific for YIG layers grown by laser MBE and do not necessarily apply to the other growth techniques such as liquid phase epitaxy.

## II. EXPERIMENTAL

The epitaxial YIG layers were grown at 700-1000°C by laser molecular beam epitaxy onto annealed GGG (111) substrates, following the approach addressed in our earlier works [13,14]. As described therein, growth results in high quality YIG films with sharp X-ray diffraction Bragg peaks, high contrast Laue oscillations, smooth atomically flat surface, ultra-narrow magnetization loops and low spin waves damping coefficient [6]. The surface morphology characterization by atomic force microscopy (Fig. 1) showed that YIG layers are atomically flat, exhibiting the step-and-terrace surface morphology that is typical of the layer-by-layer growth. The well-pronounced high energy electron diffraction (RHEED) intensity oscillations (Fig. 1 (b)) were observed during film deposition confirming the layer-by-layer growth and allowing precise calibration of the growth rate and film thickness.

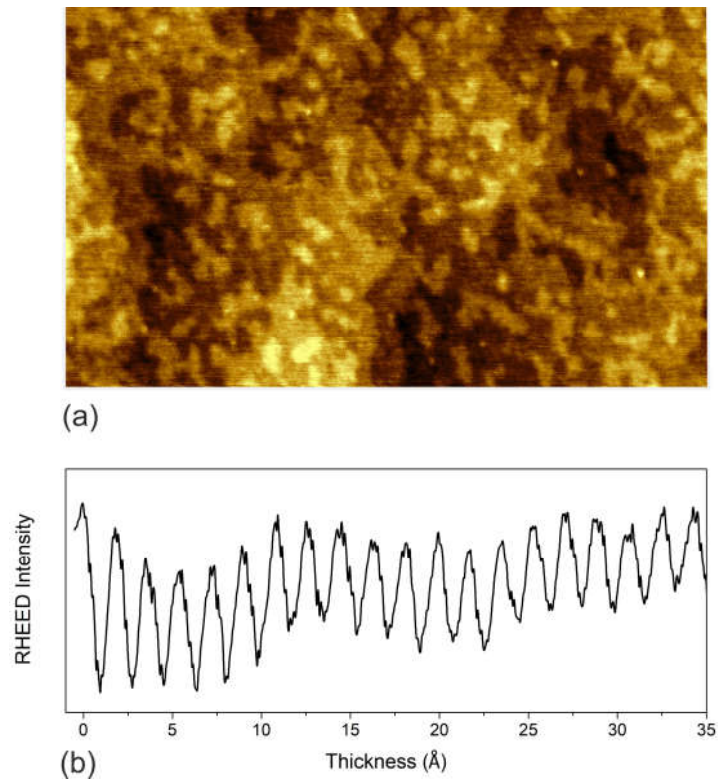

FIG. 1. The AFM image of the step-and-terrace surface morphology observed in 20 nm thick YIG layer grown at 850°C (a). The AFM image size is 600 nm × 1000 nm × 1 nm. The RHEED intensity oscillations are observed during the layer-by-layer YIG growth (b).

The X-ray Photoelectron Spectroscopy (XPS) measurements were carried out to study chemical composition and oxidation state in the near surface region. A Physical Electronics 15-255G AR double pass CMA electron energy analyzer and a double anode XR3 X-ray source (VG Microtech) operated at 15 kV, 18 mA for Mg K$\alpha$ photons were used. No surface sputtering was performed prior to XPS studies as the latter is known to change oxidation state of Fe. Secondary ion mass spectroscopy (SIMS) was applied for chemical composition depth profiling. The mass-spectra of positive secondary ions were collected from a 150 mkm area using 25 kV Bi ions. Sputtering was performed applying 1 kV Ar ions in a 200 mkm crater.

PNR was applied to probe depth dependent nuclear and magnetic scattering length densities. The measurements were performed at the Super ADAM setup [21] (Institut Laue-Langevin, Grenoble, France) with a monochromatic beam (wavelength $\lambda$ = 5.18 Å) and polarization P = 99.8 %. The neutron reflectivity for polarizations parallel ($R^+$) and antiparallel ($R^-$) to the in-plane magnetic field of 500 Oe was measured as a function of temperature. X-ray reflectivity (XRR) was used complementary to PNR to get information on the electronic density depth profiles. The reflectivity curves were measured at a wavelength of $\lambda$ = 1.04 Å at BL3A beamline of Photon Factory synchrotron (Tsukuba, Japan). The fitting of PNR and XRR was performed using the GenX package [22].

The high frequency magnetic response of the YIG films was measured by ferromagnetic resonance (FMR) and spin wave propagation spectroscopies to get complementary information on the standing and travelling spin waves. FMR spectra were measured with a conventional electronic paramagnetic resonance (EPR) spectrometer at the fixed microwave frequency of 9.4 GHz. The spin wave propagation was studied in the Damon-Eshbach setup [23]. The YIG/GGG samples were placed on the microstripe antennas with 30 μm thickness, 2 mm length, and 1.2 mm separation. The transmission coefficient $S_{21}$ was measured with the Rohde-Schwarz ZVA-40 vector network analyzer in the fixed magnetic field of 550-650 Oe applied in-plane.

The choice of YIG film thickness was guided by the need to distinguish the modified interface from the main YIG layer. For the depth resolving methods such as SIMS, PNR and XRR, the total film thickness (16-20 nm) was chosen to significantly exceed the dead layer thickness (few nm). With those techniques for which depth sensitivity was not available, we studied thickness series of 4 – 6 – 15 - 25 nm by FMR and SW spectroscopies and 4 - 13 nm by XPS. The main results in this paper were obtained for the films grown at 700-850°C. At a higher growth temperature of 1000°C it was difficult to keep stoichiometry (as shown in the XPS section below). Neither did we go lower than 700°C, as in this case the crystalline quality deterioration needs compensation by post growth annealing.

**III. QUENCHING OF HIGH FREQUENCY DYNAMIC MAGNETIC PROPERTIES IN ULTRATHIN YIG FILMS**

As it was demonstrated in Ref. [16] by measuring M(H) loops of YIG films with different thicknesses, the saturation magnetization decreases linearly with the decrease of the film thickness and approaches zero at the film thickness of 6-7 nm. This indicates that a magnetically dead layer is present at the YIG / GGG interface from the point of view of static magnetometry. Taking into account that the high frequency magnetic response of YIG layers is very important in spintonic applications, we have performed a similar study with respect to the dynamic magnetic properties. We have investigated the thickness dependence of the ferromagnetic resonance and spin wave propagation.

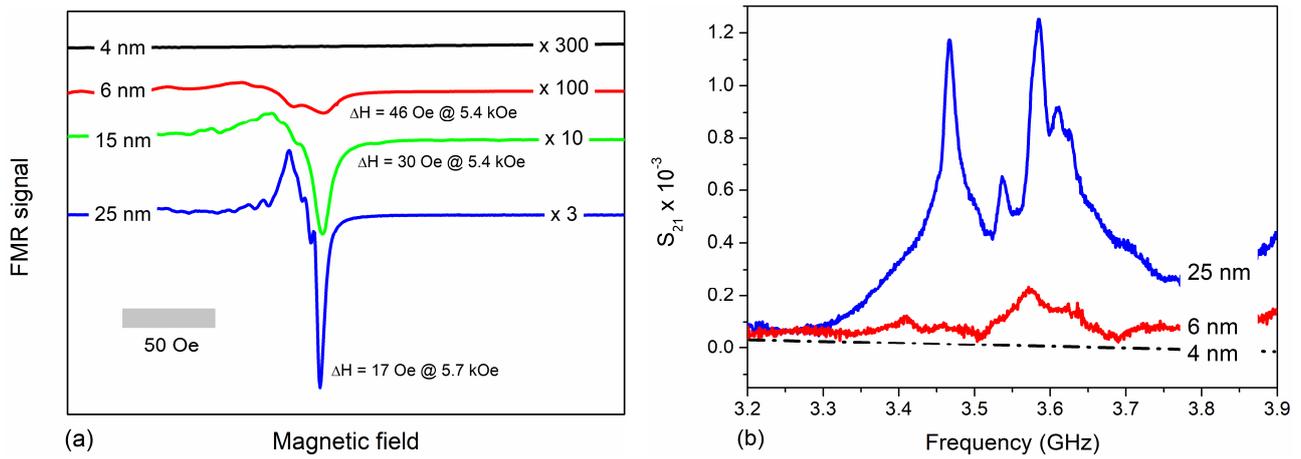

FIG. 2. The weakening of the high-frequency magnetic response in the ultrathin YIG films as demonstrated by FMR (a) and spin wave transmission (b) spectra in a series of YIG layers with different thickness. The FMR spectra are measured in the magnetic field perpendicular to the film plane. The spectra are scaled, shifted vertically and aligned horizontally for ease of comparison.

Fig. 2 (a) shows the FMR spectra measured in a series of YIG layers of different thicknesses. One can clearly observe that while the film thickness is decreasing from 25 nm to 6 nm, the FMR line is getting wider and lower in intensity. The resonance is still detectable but extremely weak in the 6 nm YIG film, and no FMR signal can be found in 4 nm film. A similar behavior of the FMR linewidth as a function of film thickness was reported earlier by Sun et al. [7]. Spin wave propagation shows the same trends: as it is shown in Fig. 2 (b) for the same thickness series of YIG samples, the spin wave transmission coefficient $S_{21}$ drastically decreases with the decrease of the film thickness from 25 nm to 6 nm. The response of the 4 nm film is only traced schematically as a flat line in Fig. 2 (b), the signal measured did not emerge above the noise level for this film. The shape of the spin wave transmission spectrum with additional peaks at lower values of effective magnetization $4\pi M - H_a$ (where $4\pi M$ is the magnetization and $H_a$ is the uniaxial anisotropy field) can be a result of depth inhomogeneity, indicating the presence of a transition layer between the GGG substrate and the YIG film. The highest transmission coefficient of -29 dB was observed in the 25 nm YIG film.

## IV. DEPTH RESOLVED CHEMICAL COMPOSITION OF YIG / GGG FILMS BY SIMS AND XPS

The quenching of static and dynamic magnetic properties in the ultrathin YIG layers suggests that a magnetically dead layer exists at the YIG / GGG interface. In order to shed light on the origin of this layer, two complementary methods - SIMS and XPS - have been applied in the present work to study the depth dependent chemical compositions of the YIG / GGG layers. SIMS was used to obtain element-selective depth profiles, without quantitative evaluation of the element concentrations. XPS was used to non-destructively obtain the chemical composition of the YIG film near surface region and to monitor the iron oxidation states. The SIMS profiles of Fe, Y, Ga and Gd measured in YIG layers grown at 700°C and 850°C are shown in Fig. 3. The profiles are corrected to give flat 100 % concentration of Ga and Gd deep inside the GGG substrate. The gray rectangle marks an approximately 7 nm thick region, where the profiles show similar broadening, due to film inhomogeneity. The features present in this region cannot be easily interpreted as they correspond to a convolution of concentration, substrate roughness, film inhomogeneity and change of the ionization efficiency at the interface.

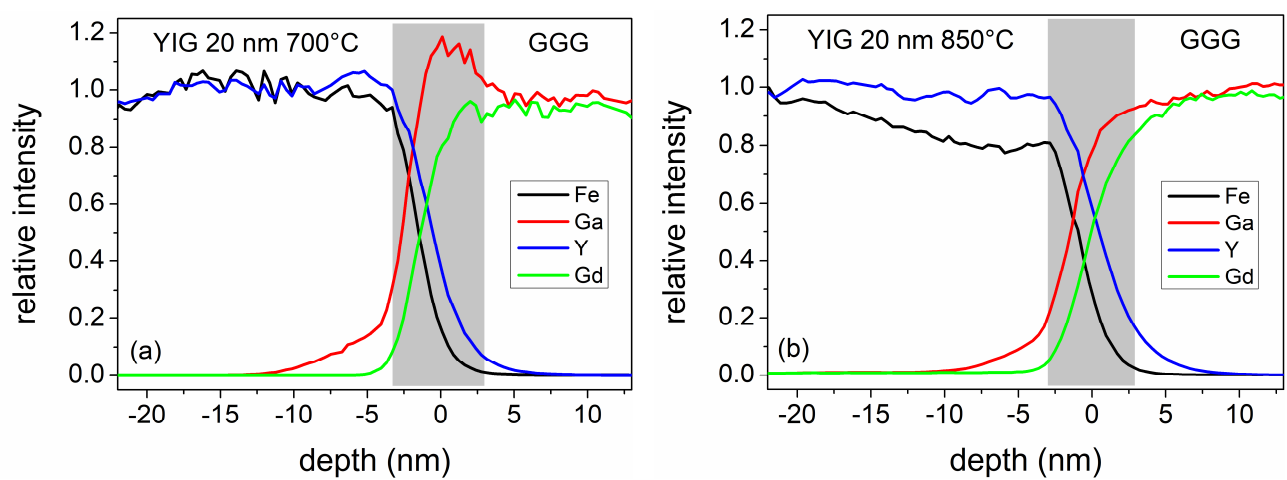

FIG. 3. SIMS profiles of Fe, Y, Ga and Gd positive ions measured in the YIG layers grown at 700°C (a) and 850°C (b). The profiles illustrate noticeable Ga diffusion into the YIG film interfacial region and variation of Fe:Y ratio in the 850°C film.

Interestingly, in all the studied samples the Ga concentration profiles extend up to 5-7 nm deep into the YIG film while the Gd profile sharply drops to zero beyond the gray-labeled broadening region. This observation suggests that diffusion of Ga atoms into the film occurs during the growth. The Fe:Y ratio noticeably decreases towards the interface in the samples grown at 850°C and stays almost constant in the YIG film grown at 700°C. Thus, we believe that during the high temperature growth stage the iron atoms in YIG are partially substituted with gallium atoms that penetrate into the YIG film from the GGG substrate to the depth of several nanometers. The back diffusion of Fe and Y into the substrate, if any, is difficult to estimate accurately, as the border between the "gray" transition layer and the substrate is not well defined. While Y tends to extend farther into GGG than Fe, it can be also due to an instrumental effect. The observed Ga/Fe concentration profile behavior resembles that reported by Ukleev et al. [24] for the εFe$_2$O$_3$ / GaN system, where the partial substitution of iron by gallium in the interface region was demonstrated by SIMS.

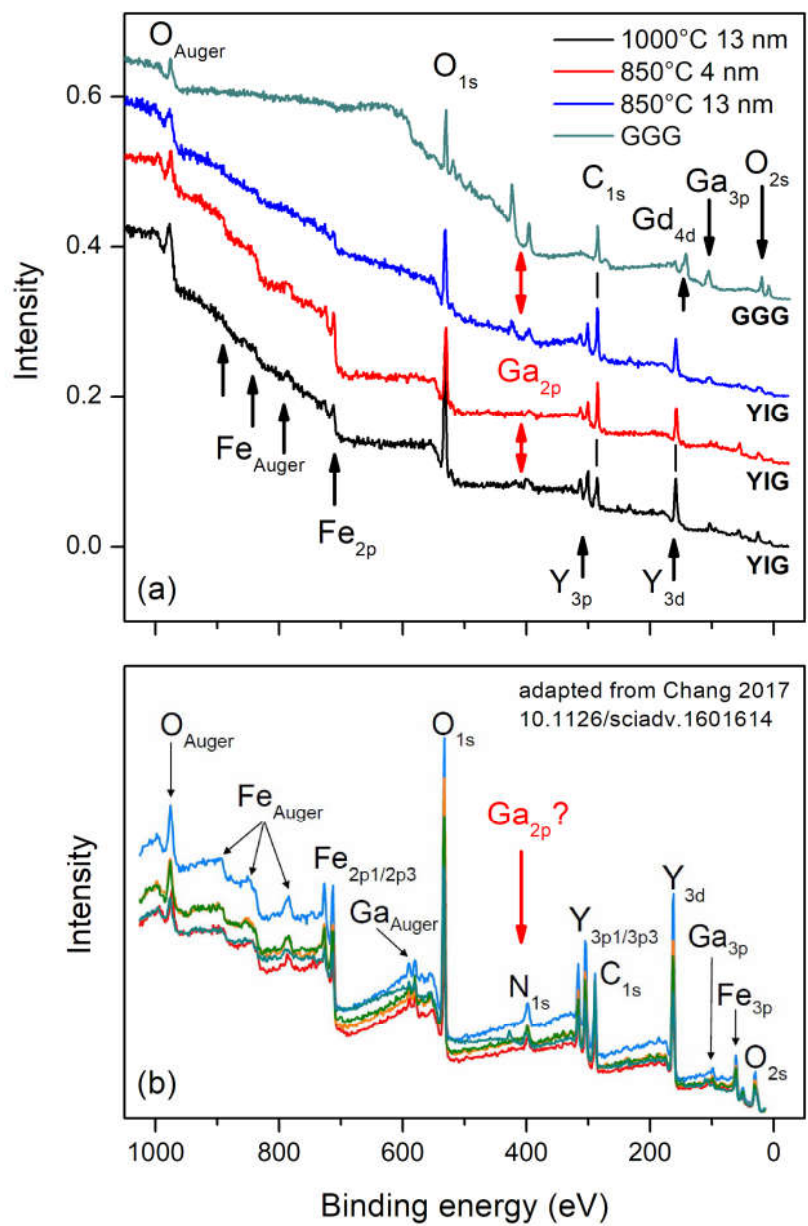

FIG. 4. Wide energy range XPS spectra measured for GGG substrate and a series of YIG / GGG layers grown at 850°C – 1000°C (a). Shown for comparison are the XPS spectra measured for sputter annealed 20 nm YIG / GGG layers adapted from [25] (b).

The wide range X-ray photoemission spectra measured in 13 nm films grown at 850°C and 1000°C, 4 nm film grown at 850°C and clean GGG(111) substrate are shown in Fig. 4. Characteristic photoemission and Auger peaks of Fe, Y, O (YIG film), Ga, Gd (GGG substrate) and C, N (post growth contaminants) show up, confirming the chemical pureness of the YIG films. The Fe 2p, Y 3p, Y 3d and O 1s photoemission spectra are shown in higher resolution in Fig. 5. The levels of yttrium and oxygen are not much varied in the studied samples. The amount of Fe on the surface is the lowest in the 4 nm sample grown at 850°C, is slightly higher in the 13 nm sample grown at 850°C, and is the highest in the 13 nm sample grown at 1000°C. In the similar study of the PLD grown YIG/GGG films [7], the Y:O ratio was claimed to be constant (3:12), while the Y:Fe ratio was shown to vary from 3:2.1 to 3:2.6 corresponding to iron deficiency (should be 3:5 in stoichiometric $Y_3Fe_5O_{12}$). The Fe content was stated to increase with the growth of the temperature. Our finding is consistent with this observation.

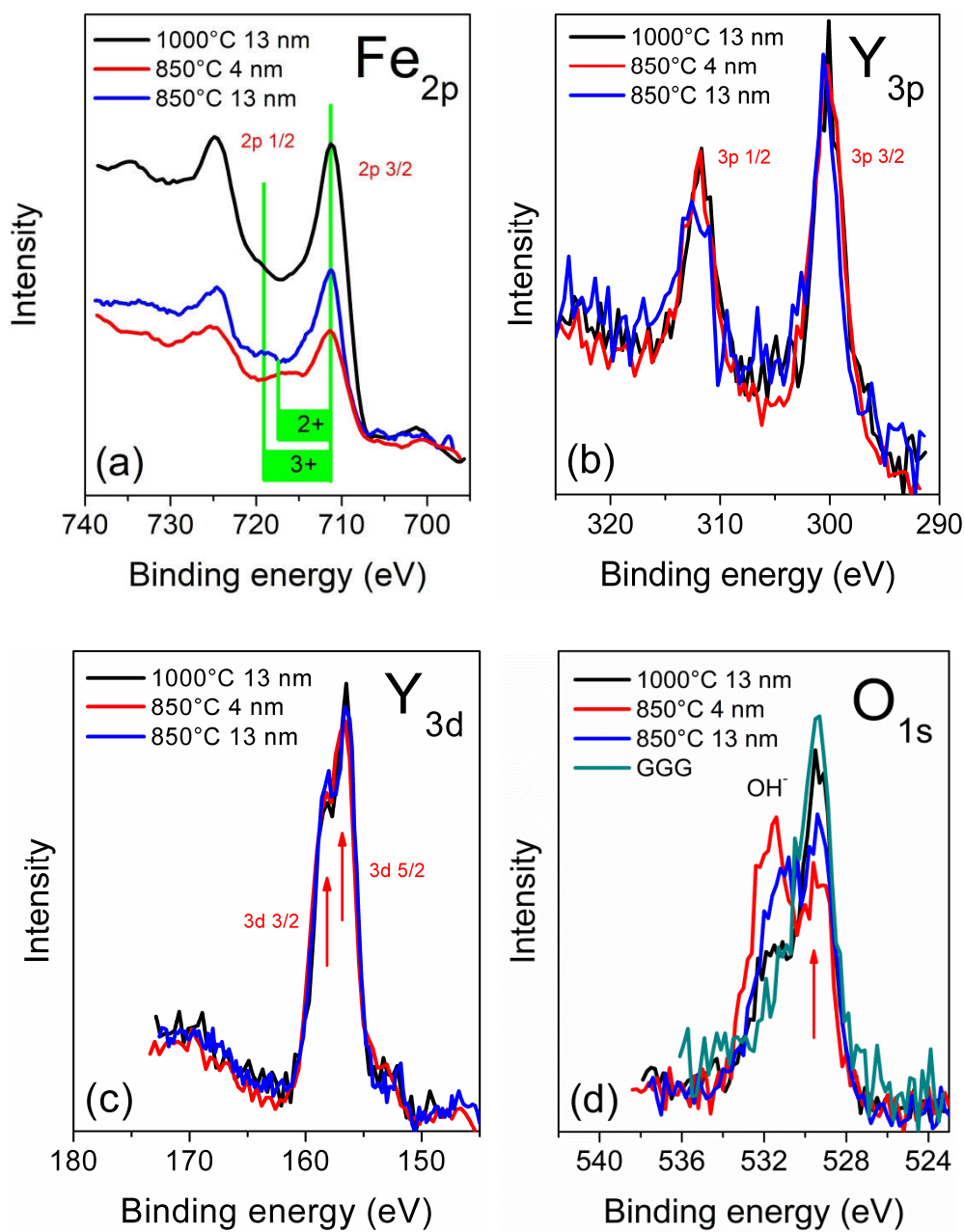

FIG. 5. XPS spectra of Fe 2p, Y 3p, Y 3d and O 1s measured in a series of YIG films grown on GGG at 850°C – 1000°C. To facilitate the comparison of the spectral shapes and intensities, the background is subtracted, and the normalization to yttrium is performed for every spectrum.

As it is shown in Fig. 5, the Fe $2p_{3/2}$ and $2p_{1/2}$ peaks are positioned at 710.9 eV and 724.5 eV, respectively, independent on the growth conditions. No indication of metallic Fe at 707 eV [26] is present. The positions of Fe 2p peaks are often used to estimate the presence of Fe 2+ / 3+ valence mixing. For example, in Ref. [27] such mixing was claimed to exist in Bi substituted YIG films grown by PLD. Similar considerations were given in Ref. [28] for Zr doped YIG, in Ref. [29] for Ce doped YIG and in Ref. [30] for Bi doped YIG films obtained by magnetron sputtering. Due to the existing ambiguities regarding the absolute positions of $Fe^{2+}$ and $Fe^{3+}$ peak maxima, it is more appropriate to distinguish $Fe^{3+}$ from mixed $Fe^{3+}$ / $Fe^{2+}$ by the satellite shoulder that appears at high binding energy side of the $2p_{3/2}$ peak at a distance of ~8 eV for pure 3+ and ~6 eV for pure 2+. In the mixed valence compounds such as $Fe_3O_4$, the satellite is usually not prominent due to the superposition of $Fe^{2+}$ and $Fe^{3+}$ associated satellites. In the XPS studies related to thick YIG films, the 3+ satellite is usually observed [7,25,29,31] indicating the domination of $Fe^{3+}$. In the Fe 2p spectra shown in Fig. 5a, the 4 nm film grown at 850°C exhibits a trace of 2+ satellite, while 13 nm films grown at 850°C and 1000°C films show a signature of 3+ satellite. Thus, it is likely that the $Fe^{3+}$ state characteristic for bulk YIG is mostly obtained in thick YIG films, while in the thin YIG film there is a trace of 2+ iron. As it was shown in [32] the presence of Fe 2+ ions in thin YIG film can lead to significant increase of the FMR linewidth at low temperature.

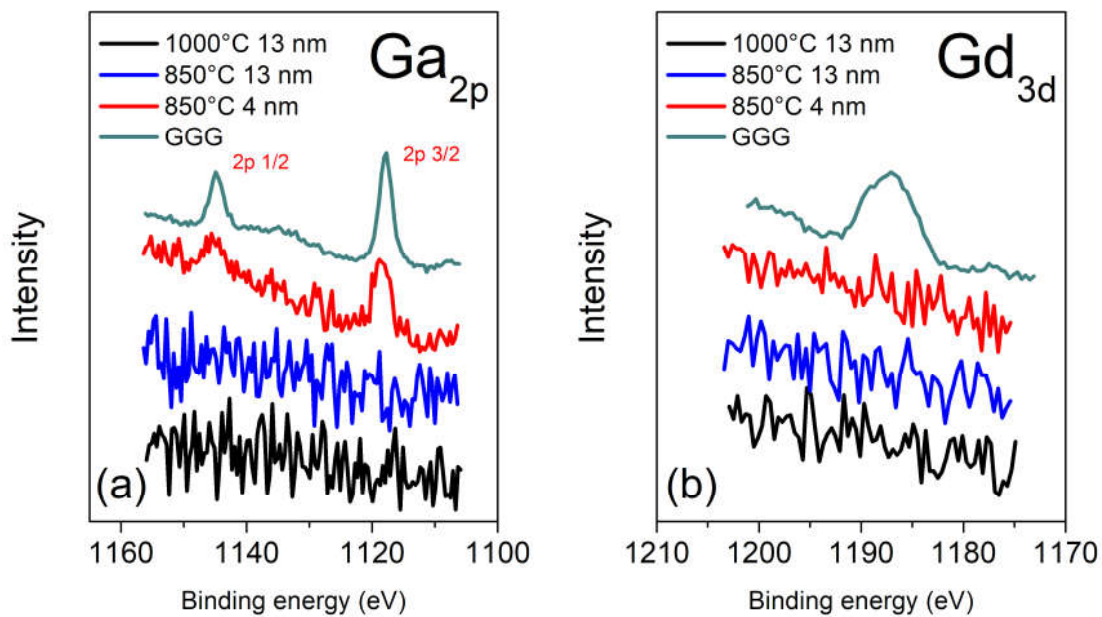

FIG. 6. XPS spectra of Ga 2p (a) and Gd 3d (b) measured in the GGG substrate and in YIG / GGG films grown at 850°C – 1000°C.

The O 1s spectra in Fig. 5 (d) show two peaks at binding energies of 529.5 eV and 531.6 eV.

Of these two, the peak at lower binding energy (which is the only one observed in the GGG substrate) corresponds well to the position reported for $Fe_2O_3$, $Y_2O_3$ and $Y_3Fe_5O_{12}$ [31,33,34]. The second peak on the high binding energy side is supposed to be related to the hydroxile group due to surface contamination [31,35].

Interestingly, the XPS spectrum of the 4 nm YIG film grown at 850°C shows a noticeable Ga 2p peak (Fig. 6 (a)). At the excitation energy of the used photon (1253.6 eV), the kinetic energy of the photoelectrons is low, corresponding to the minimum of the mean free path for electrons in solids. Even forcing the assumption that these photoelectrons are excited in the substrate and they are only slightly attenuated by the 4 nm YIG layer, one would expect to observe also the Gd 3d peak (of comparable kinetic energy), along with the Ga 2p peak. In GGG, these two peaks have comparable intensities (see the XPS spectrum of GGG in Fig. 4 (a)) and similar attenuation depths. The fact that the Gd 3d is not observed in the 4 nm film (Fig. 6 (b)) is strong evidence that Ga is present in thin YIG layer. From the absence of the Ga 2p peak in the 13 nm films, we conclude that Ga is only present in the few nanometer thick interface region of the YIG film. The slight gallium concentration increase in the vicinity of the YIG/GGG interface was previously reported by XPS [19]. Gallium signature was also reported in Ref. [25], where YIG layers of 20 nm thickness were grown on GGG (111) by magnetron sputtering at RT followed by few hours annealing in oxygen at 800°C. The XPS spectra presented in Ref. [25] show noticeable traces of Ga (both Auger and PE peaks) that are, however, claimed by the authors to come from the GGG substrate. This is arguable as the photoelectrons excited in the GGG substrate are not supposed to be able to escape through a 20 nm thick YIG film. Moreover, if Ga photoemission from the substrate is visible, so would be the Gd photoemission, especially at low binding energies for which electron kinetic energies are high. However, the Gd 4d peak at 142 eV is not present in these spectra (see Fig. 4). This leads to a conclusion that in the YIG films discussed in [25] Ga diffusion from the substrate might be also present.

**V. COMPOSITION AND MAGNETIZATION IN-DEPTH PROFILING BY PNR AND XRR**

To further analyze the composition and magnetization depth profiles in the YIG / GGG system, we have applied polarized neutron and X-ray reflectometry techniques. The joint use of both methods gives advantage of complementary information: nuclear $\rho_n$ and magnetic $\rho_m$ scattering length densities (SLD) by PNR and electron $\rho_e$ density by XRR. Despite the rather moderate contrast of the real parts of $\rho_n$ and $\rho_e$ in YIG/GGG, gadolinium can be reliably distinguished from the other elements by the noticeable contribution to the imaginary part of the nuclear SLD. In our study, the reflectivity measurements were carried out for two samples: 23 nm YIG film grown at 700°C was studied by PNR and 16 nm film grown at 850°C was studied by XRR.

For quantitative discussions of nuclear, magnetic and electronic density distribution across the heterostructure we performed fitting using the Parratt algorithm [36] in GenX software package [22]. The fitting was performed using the simplest possible model. To avoid stagnation of the algorithm to a local minimum, the PNR fit was performed simultaneous for all the experimental data - all the structural parameters were kept constant between the PNR curves measured at different temperatures and magnetic fields, while the magnetizations of layers were varied. In what follows we plot the output of the fitting algorithm as the depth profiles of the complex density: nuclear and magnetic for PNR or electronic for XRR.

The experimental and fitted PNR curves measured at $T$ = 300 K, 50 K and 5 K for 23 nm YIG sample are shown in Fig. 7 (a). The reasonable fitting was obtained with three-layer model containing the substrate, transition layer, and main YIG layer. The nuclear SLDs of the GGG substrate and main YIG layer were fixed to the bulk densities. The transition layer was modeled by $Gd_xY_{3-x}Ga_yFe_{5-y}O_{12}$ chemical composition assuming the gradual substitution of Gd atoms by Y atoms and Ga atoms by Fe atoms. The depth-resolved structural and magnetic SLD profiles delivered by fitting are shown in Fig. 7 (b). Interestingly, the resultant profiles of the real and imaginary parts of the $\rho_n$ show a drastically different behavior. Due to the noticeable neutron absorption of Gd nuclei, the imaginary part of $\rho_n$ is proportional to Gd concentration. According to our fitted model Im($\rho_n$) drops sharply at $z$ = 0 indicating sharpness of the Gd profile. The observed Gd gradient is in agreement with the typical values of the GGG substrate surface roughness [13]. No noticeable diffusion of Gd atoms into the YIG film is observed. At the same time, the real part of SLD exhibits a smooth gradient extended by approximately 70 Å into the YIG film. The SLD value just above the interface (marked with a red line in Fig. 7 (b)) fits well to the SLD of $Y_3Ga_5O_{12}$ compound produced by the substitution of Fe atoms in YIG on Ga, which is in agreement with the SIMS and XPS data described above. The composition of the transition layer can be reasonably modeled by the $Y_3Ga_yFe_{5-y}O_{12}$ formula that corresponds to the gradual transition from $Y_3Ga_5O_{12}$ to $Y_3Fe_5O_{12}$.

Fig. 7 (b) shows the temperature dependent magnetization depth profiles derived from the magnetic SLD $\rho_m$. The magnetic SLD $\rho_m$ is directly proportional to the magnetization component $M$ parallel to the in-plane magnetic field: $\rho_m$ =2.853*10$^{-9}$*$M$ Å$^{-2}$, where $M$ is given in emu/cm$^3$ units.

The magnetization changes almost synchronously with the real part of the nuclear SLD $\rho_n$. At room temperature the magnetization of the main YIG layer equals 105 emu/cc, which is slightly lower than the saturation magnetization $M_s$=140 emu/cc of the bulk YIG [37] but comparable to room temperature $M_s$ values discussed in Ref. [16]. The magnetization of the transition layer drops gradually towards the YIG/GGG interface synchronously with the Ga - Fe substitution. In contrast to the recent works [16,38], we have not observed any antiparallel magnetic moment at the interfacial region at low temperatures. The different magnetic properties of the interface region could be the result of different way of YIG film preparation: while we grow the film in one stage, the authors of Refs. [16,38] use a 2 h post growth annealing. According to our experiment, the small parallel magnetic moment is observed in the interface layer at 5-50 K. The magnetization of the main YIG layer increases drastically as the temperature decreases to 5 K in agreement with Refs. [37] and [16].

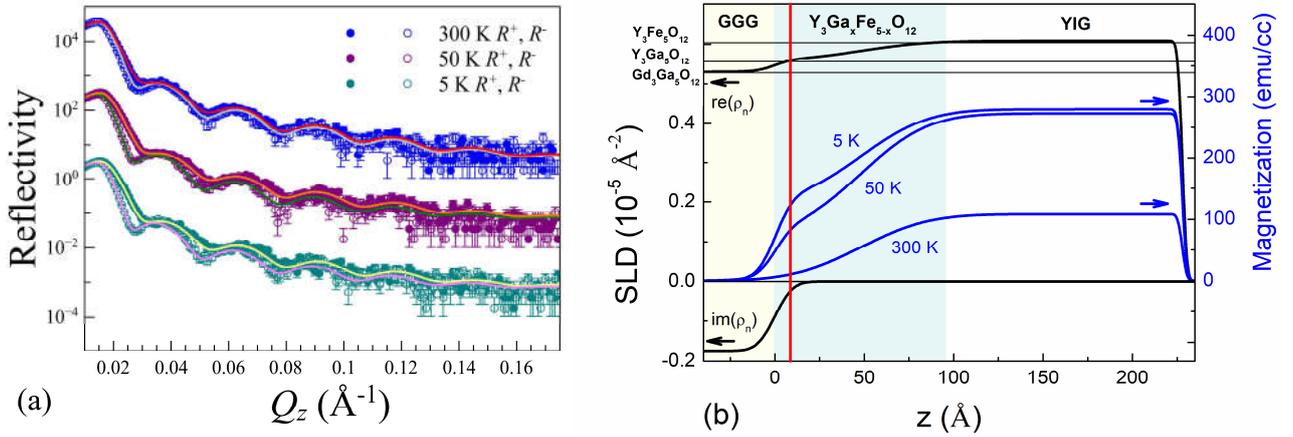

FIG. 7. (a) PNR curves of 23 nm $Y_3Fe_5O_{12}$ film grown at 700 °C measured at the applied magnetic field H = 500 Oe at T= 300 K, 50 K and 5 K. Symbols correspond to the experimental data points, while the solid lines show the fitted curves. (b) Nuclear and magnetic SLD profiles of the YIG/GGG heterostructure are obtained from the fitting routine. The bulk SLD values of the compounds are shown with the horizontal lines.

The fitted X-ray reflectivity curve measured in 16 nm $Y_3Fe_5O_{12}$ is shown in Fig. 8 (a). The reasonable fit was achieved for a four-layered model containing the GGG substrate, transition $Y_3Ga_xFe_{5-x}O_{12}$ layer, main $Y_3Fe_5O_{12}$ layer, and the peculiar low density top layer. The electron densities of the GGG substrate and main YIG layer were fixed to the corresponding bulk values. The resultant SLD profile in Fig. 8 (b) exhibits a sharp drop at z=0 Å, followed by a slow density decrease towards z=70 Å. At this point the density becomes equal to the SLD value of YIG bulk. Following the same strategy as in the PNR section above, we assume that the double slope observed is the superposition of the sharp depth profile of Gd, and the sloping profile of Ga distribution, expanding by the diffusion into the YIG layer. This assumption is in quantitative agreement with the density profile as the SLD value just above the interface (marked with the red line in Fig. 8 (b)) fits well to the electron density of $Y_3Ga_5O_{12}$ compound produced by substituting all Fe atoms in YIG by Ga. The rest of the slope can be modeled assuming the gradual transition from the compound of $Y_3Ga_5O_{12}$ to the compound of $Y_3Fe_5O_{12}$. The $Y_3Ga_xFe_{5-x}O_{12}$ composition of the transition layer with x changing from 5 down to 0 on the length scale 50-70 Å correlates well with the Ga diffusion discussed above.

Interestingly, the obtained SLD profile suggests that 15 Å top layer with reduced density exists on the top of the YIG surface. Without this layer it is impossible to model the low frequency oscillations in the reflectivity curve having maximum at $Q_z=0.4$ Å$^{-1}$ in Fig. 8 (a). The similar feature attributed to the oxidation or contamination was also present in the XRR data in Ref. [39]. The low density layer residing at the YIG surface is considered as $Y_2O_3$ in the PNR study of Cooper et al [17]. We, however, believe that this is not the case as the SLD of $Y_2O_3$ ($3.75\times10^{-5}$ Å$^{-2}$) is significantly larger than the density value observed in our work and in [17]. We think that the low density layer is rather related to the particularly-organized step-and-terrace structure of the YIG surface. As evidenced by AFM studies [6,13,14], the YIG surface is terminated by a multistoried structure consisting of (111) monolayers. When the level occupancy is uniformly decreasing towards vacuum, XRR can be modeled by the Gaussian surface roughness (Fig. 8 (c)). If the level occupancy distribution function shows a jump, the density profile will show a jump as well (Fig. 8 (d)).

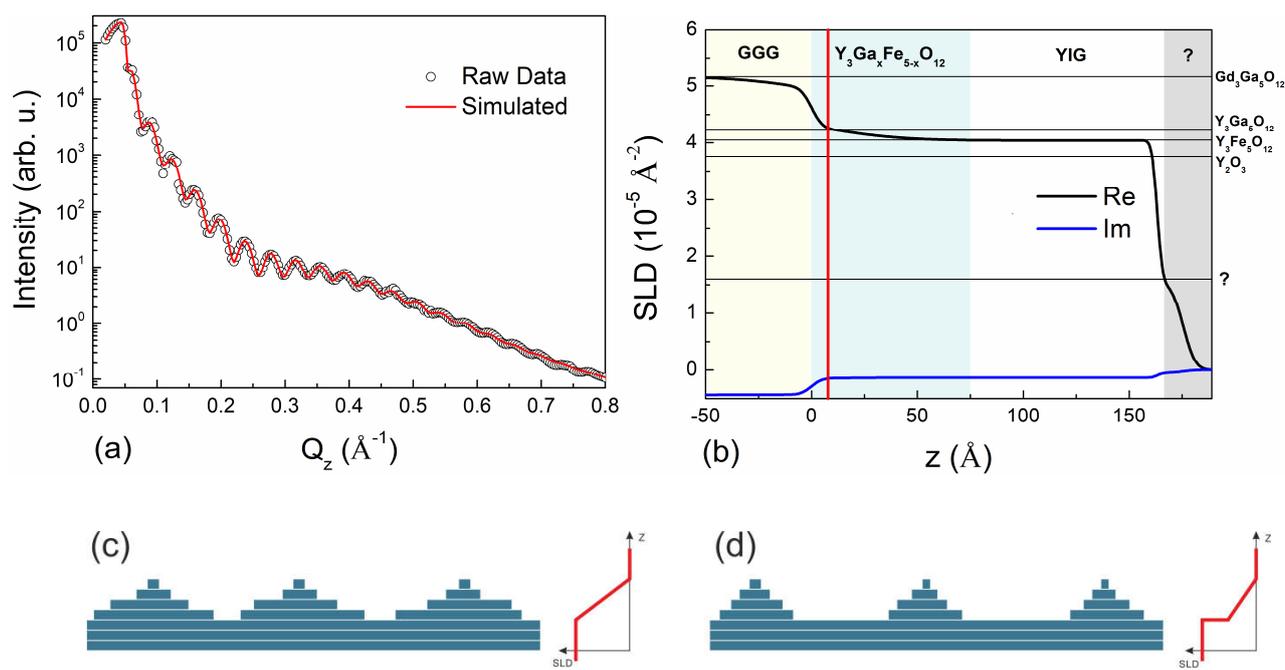

FIG. 8. X-ray reflectivity of 16 nm $Y_3Fe_5O_{12}$ film grown at 850 °C (a). Circles represent raw data, and

*solid curve is the GenX fitting. Real and imaginary parts of SLD profile for YIG film corresponding to the best fit (b). The bulk SLD values of the involved compounds are shown with the horizontal lines. The low density top layer is shown in gray and its SLD value is labeled with a question mark symbol. The sketch of the step-and-terrace surface morphologies gives uniformly sloped (c) and stepped (d) density profiles.*

The uneven level occupancy might be caused by the adatom migration that occurs when deposition has stopped but the substrate temperature is still high. The low frequency modulations often observed around the 444 and 888 Bragg reflections provide an important evidence that the low density layer has the layered crystal structure of YIG. Noteworthy, the presented XRR and PNR data are in general agreement with each other. As the spanned $Q_z$ range accessed by PNR (0.18 Å$^{-1}$) is significantly narrow in comparison with XRR (0.8 Å$^{-1}$), the neutron reflectivity curve does not show the characteristic low frequency modulation and, therefore, provides no evidence of 15 Å thin low density layer residing on top of the YIG surface. We expect that such evidence would become available if the PNR was measured to a higher value of $Q_z$. Taking into account the data of the other groups[39] as well as our own preliminary XRD studies (to be presented elsewhere), we believe that the low density layer exists as well in the 700°C layer.

**VI. CONCLUSION**

In our paper, we shed light onto the origin of the few nanometer-thick magnetically dead layer present at the interface of the epitaxial YIG / GGG layers grown at above 700°C by means of laser MBE. Previously, the existence of such layer was mainly suggested based on results of indirect methods, such as static magnetometry measurements. In the present work, we directly show this effect in thin YIG films by means of the ferromagnetic resonance, spin wave propagation and polarized neutron reflectometry. We have demonstrated that the resonance magnetic properties are noticeably quenched as the YIG film thickness decreases to a value of few nanometers. As opposed to the previous works [16] and [17], where the magnetically dead layer was claimed to be due to Gd diffusion, our SIMS, XPS, XRR and PNR measurements have shown no trace of Gd migration into the YIG layer. We have revealed 5-7 nm interface region in the YIG layer – Ga-rich and deficient of Fe. The Ga diffusion was further confirmed by the reflectivity measurements performed by polarized neutrons and X-rays. The PNR study has shown that the magnetization within the dead layer gradually decreases from the quasi bulk value in the main YIG layer to zero at the interface. The magnetization was shown to increase by the factor of 2.5 at low temperature. The small non-proportional increase of magnetization within the interface layer was observed upon the sample cooling below 50 K, possibly due to the magnetic phase transition. To our knowledge, this is the first time that Ga diffusion during YIG / GGG growth is directly confirmed by a combination of direct space and reciprocal space methods. Interestingly, a peculiar 15 Å low density layer residing on top of the YIG layer has been observed by the X-ray reflectivity. In our opinion, this layer cannot be explained by a uniform film of a crystalline phase, as it was claimed by [17] regarding $Y_2O_3$. We do rather suggest that the peculiar low frequency oscillations in the XRR reflectivity curve are caused by a non-uniform height distribution within the step-and-terrace multilevel structure at the YIG surface. The presented SIMS, XPS, PNR and XRR data provide strong evidence that the YIG/GGG interface region is magnetically different from the YIG bulk due to the intermixing caused by Ga diffusion from the GGG substrate. It must be noted that the presented results are specific for the YIG layers grown in the 700 – 850°C temperature range in which the YIG/GGG interface peculiarities do not show drastic temperature dependence (with somewhat flatter concentration profiles for 700°C films). The results are obtained for films grown by laser MBE and do not necessarily apply to the other YIG growth techniques such as liquid phase epitaxy or magnetron sputtering.

**ACKNOWLEDGMENTS**

The authors gratefully acknowledge the fruitful discussions with B.B. Krichevtsov, as well as the collaborative research involving Photon Factory synchrotron (proposals 2014G726 and 2016G684) and ILL (proposal CRG-1992) facilities. We thank O. Aguettaz (ILL) for technical assistance. The authors wish to acknowledge the beamline staff at PF for kind assistance in conducting experiments.

**FUNDING INFORMATION**

The work was funded by Russian Science Foundation (project 17-12-01508). The part of the study related to PNR was supported by Sinergia CDSII5-171003 NanoSkyrmionics.